\documentclass[12pt,twocolumn,preprint,trackchanges]{aastex62}
\usepackage{natbib}
\usepackage{multirow,color}
\usepackage{bbding}
\usepackage{amssymb}
\usepackage{ulem}
\usepackage{color}
\usepackage{longtable}
\usepackage{array}
\usepackage{bm}
\usepackage{url}

\def\gsim{\;\lower4pt\hbox{${\buildrel\displaystyle >\over\sim}$}\;}
\def\lsim{\;\lower4pt\hbox{${\buildrel\displaystyle <\over\sim}$}\;}
\def\grls{\;\lower4pt\hbox{${\buildrel\displaystyle >\over <}$}\;}

\bibpunct{(}{)}{,}{o}{}{,} 

\begin{document}
\title{Rapid buildup of a magnetic flux rope during a confined X2.2 class flare in NOAA AR 12673}

\author{Lijuan Liu}
\affiliation{School of Atmospheric Sciences, Sun Yat-sen University, Zhuhai, Guangdong, 519082, China}
\email{liulj8@mail.sysu.edu.cn}

\author{Xin Cheng}
\affiliation{School of Astronomy and Space Science, Nanjing University, Nanjing, 210023, China}
\email{xincheng@nju.edu.cn}

\author{Yuming Wang}
\affiliation{CAS Key Laboratory of Geospace Environment, Department of Geophysics and Planetary Sciences, University of Science and Technology of China, Hefei, Anhui, 230026, China}

\author{Zhenjun Zhou}
\affiliation{School of Atmospheric Sciences, Sun Yat-sen University, Zhuhai, Guangdong, 519082, China}

\author{Yang Guo}
\affiliation{School of Astronomy and Space Science, Nanjing University, Nanjing, 210023, China}

\author{Jun Cui}
\affiliation{School of Atmospheric Sciences, Sun Yat-sen University, Zhuhai, Guangdong, 519082, China}
\affiliation{CAS Key Laboratory of Lunar and Deep Space Exploration, National Astronomical Observatories, Chinese Academy of Sciences, Beijing, 100012, China}



\begin{abstract}

Magnetic flux ropes (MFRs) are believed to be the core structure in solar eruptions, 
nevertheless, their formation remains intensely debated. 
Here we report a rapid buildup process of an MFR-system during a confined X2.2 class flare occurred on 2017 September 6 in NOAA AR 12673, 
three hours after which the structure erupted to a major coronal mass ejection (CME) accompanied by an X9.3 class flare. 
For the X2.2 flare, 
we do not find EUV dimmings, separation of its flare ribbons, or clear CME signatures, 
suggesting a confined flare. 
For the X9.3 flare, large-scale dimmings, 
separation of its flare ribbons, 
and a CME show it to be eruptive.  
By performing a time sequence of nonlinear force-free fields (NLFFFs) extrapolations we find that: until the eruptive flare, an MFR-system was located in the AR.  
During the confined flare,  
the axial flux and the lower bound of the magnetic helicity for 
the MFR-system were dramatically enhanced by about $86\%$ and $260\%$, respectively, 
although the mean twist number was almost unchanged. 
During the eruptive flare, the three parameters were all significantly reduced. 
The results evidence the buildup and release of the MFR-system during the confined and the eruptive flare, respectively. 
The former may be achieved by flare reconnection. 
We also calculate the pre-flare distributions of the decay index 
above the main polarity inversion line (PIL) 
and find no significant difference. 
It indicates that the buildup of the magnetic flux and helicity of the MFR-system 
may play a role in facilitating its final eruption. 

\end{abstract}

\section{INTRODUCTION}\label{sec:intro}

Magnetic flux ropes, 
consisting of twisted field lines winding around the main axis, are fundamental structures on the sun. 
It's generally accepted that the main drivers of hazardous space weather - CMEs - are expulsions of MFRs~\citep[e.g.,][]{Amari_2000, Chengx_2017}. 
Therefore, their properties, e.g., formation, evolution, stability, are long-standing topics. 
In most solar eruption models, the MFRs play key roles. 
In some ideal magnetohydrodynamic (MHD) models, 
pre-existing MFRs are required: 
once the background magnetic fields decrease fast enough~\citep[torus instability,][]{Kliem_2006}, or the rope's twist number reaches a critical value \citep[kink instability,][]{Torok_2004}, 
the eruption happens. 
In some non-ideal models, MFRs can be formed during the eruption: magnetic reconnection happens below~\citep[tether-cutting model,][]{Moore_2001} or above~\citep[breakout model,][]{Antiochos_1999} the sheared magnetic arcades,  
then forms the MFR and initiates the eruption. In either kind of models, the eruption entity is the MFR.

Despite its pre-existence for an eruption, the MFR is proposed to be formed or strengthened mainly through two ways: one is its bodily emergence, 
supported by simulations~\citep[e.g.,][]{Fanyh_2001, Leakej_2013} and observations~\citep[e.g.,][]{Okamoto_2008}, though questioned by some work~\citep{Vargas_2012}; 
the other is reconnection mechanism, 
in which the MFR is formed/strengthened through the reconnection \citep[e.g.,][]{Chengx_etal_2011, Zhangj_2012, Wangws_2017}, 
sometimes manifested as flux cancellation~\citep[e.g.,][]{Green_2010}. 
The twist numbers of interplanetary MFRs are usually larger than that of pre-eruption MFRs, suggesting that the latter may be strengthened by reconnection during the 
eruptions~\citep{Wangym_2016}.
Besides, photospheric flows/motions, e.g., shear/converging motions, sunspot rotations are reported to build up the MFRs effectively~\citep[e.g.,][]{Fanyh_2009, Yanxl_2018a}. 

In the reconnection formation scenario, 
the MFRs can even be formed during confined flares before successful eruptions as reported: 
\citet{Patsourakos_2013} observed a limb event, 
showing the formation of an MFR during a confined C4.5 GOES class flare before its ejection in the subsequent M7.7 flare;   
\citet{Guoy_2013} suggested that an MFR can be built-up during a series of confined flares before a CME and a major flare;  
\citet{Chintzoglou_2015} reported formation and strengthening of two MFRs during confined flares before two successive CMEs; 
\citet{James_2018} also reported an MFR formation in a confined flare before its final eruption. 

Nevertheless, 
the formation/buildup of MFRs during as intense as X-class flares is barely reported to our best knowledge. 
Here we present a buildup process of an MFR-system during a confined X2.2 flare (SOL2017-09-06T08:57) that occurred in NOAA active region (AR) 12673, which erupted to a major CME accompanied by an X9.3 flare (SOL2017-09-06T11:53) three hours later. 
We mainly focus on analyzing the properties of the source coronal magnetic fields (CMFs) based on a time-sequence of NLFFF extrapolations.

\section{Data and Methods}\label{sec:dat_meth}

Both analyzed flares occurred near S09W33. 
Their on-disk evolution is captured by {\it SDO}/AIA~\citep{Pesnell_2012}. 
The associated coronal outflows are observed by {\it SOHO}/LASCO~\citep{Domingo_1995}.  
The photospheric vector magnetic fields are recorded by {\it SDO}/HMI. 
Here a magnetic-field data product called SHARP~\citep{Bobra_2014} is used. 
The three dimensional CMFs 
are reconstructed by an NLFFF extrapolation model 
\citep{Wiegelmann_2004, Wiegelmann_2012} and a potential fields (PFs) model~\citep[e.g.,][]{Sakurai_1989}, using SHARP magnetograms as the bottom boundaries. 

Based on the reconstructed CMFs, we may identify the MFR by combining the twist number $T_w$ and squashing factor $Q$~\citep{rliu_2016}. 
$T_w$ measures the number of turns of a field line winding, 
and is calculated by $\displaystyle T_w = \frac{1}{4\pi}\int_l\alpha dl$. $\alpha$, $l$ and $dl$ denote the force-free parameter, the field line length, and the elementary length, respectively. 
$Q$ measures the change of the connectivity of the field lines. High $Q$ values indicate the positions of quasi-separatrix layers (QSLs) where the connectivity changes dramatically \citep[e.g.,][]{Demoulin_1996}. 
For an MFR, its twisted fields lines are distinct from the surrounding fields, 
thus are wrapped by QSL. Its cross section displays a twisted region enclosed by high $Q$ lines. 

Once the MFR is located, 
its toroidal (i.e., axial) flux can be calculated by $\Phi_t=\bf{B}\cdot\bf{S}$, where ${\bf S}=S\bf{n}$, 
$S$ and $\bf{n}$ denote the area and normal unit vector of its cross section. 
The flux-weighted mean twist number can be computed by $\displaystyle <|T_w|>=\frac{\Sigma |T_wB_n|dA}{\Sigma |B_n|dA}$, 
where $dA$ is the elementary area. 
Its magnetic helicity can be simply estimated by a twist number method~\citep{Priester_2000, Guoy_2013, Yangk_2016, Guoy_2017a} 
using $H_{twist} \approx T_w\Phi^2$. Magnetic helicity measures the geometrical complexity of the magnetic fields. 
For a single MFR, the aforementioned method is applicable when its writhe can be omitted, giving the lower bound of its helicity. 

Additionally, the decay index measuring the decrease of the external 
PFs with height can be calculated by $\displaystyle n=-\frac{\rm \partial\,\ln B_{ex}(h)}{\rm \partial\,\ln h}$. 
$B_{ex}$ is the traverse component of PFs and $h$ is the height. 
As stated in torus instability theory, the background PFs play the major role in confining the eruption. 
Once the MFR reaches a region where $n$ is beyond a threshold, 
the torus instability will occur. 
The threshold value 
varies depending on the properties of the MFR \citep[e.g.,][]{Demoulin_2010, Olmedo_2010}. We use 1.5~\citep{Kliem_2006} as a representative value here.  

\section{Observation and Analysis}\label{sec:obs_ana}

\subsection{Eruptiveness of the Flares}\label{subsec:obs}

\begin{figure*}
\begin{center}
\epsscale{1.2}
\plotone{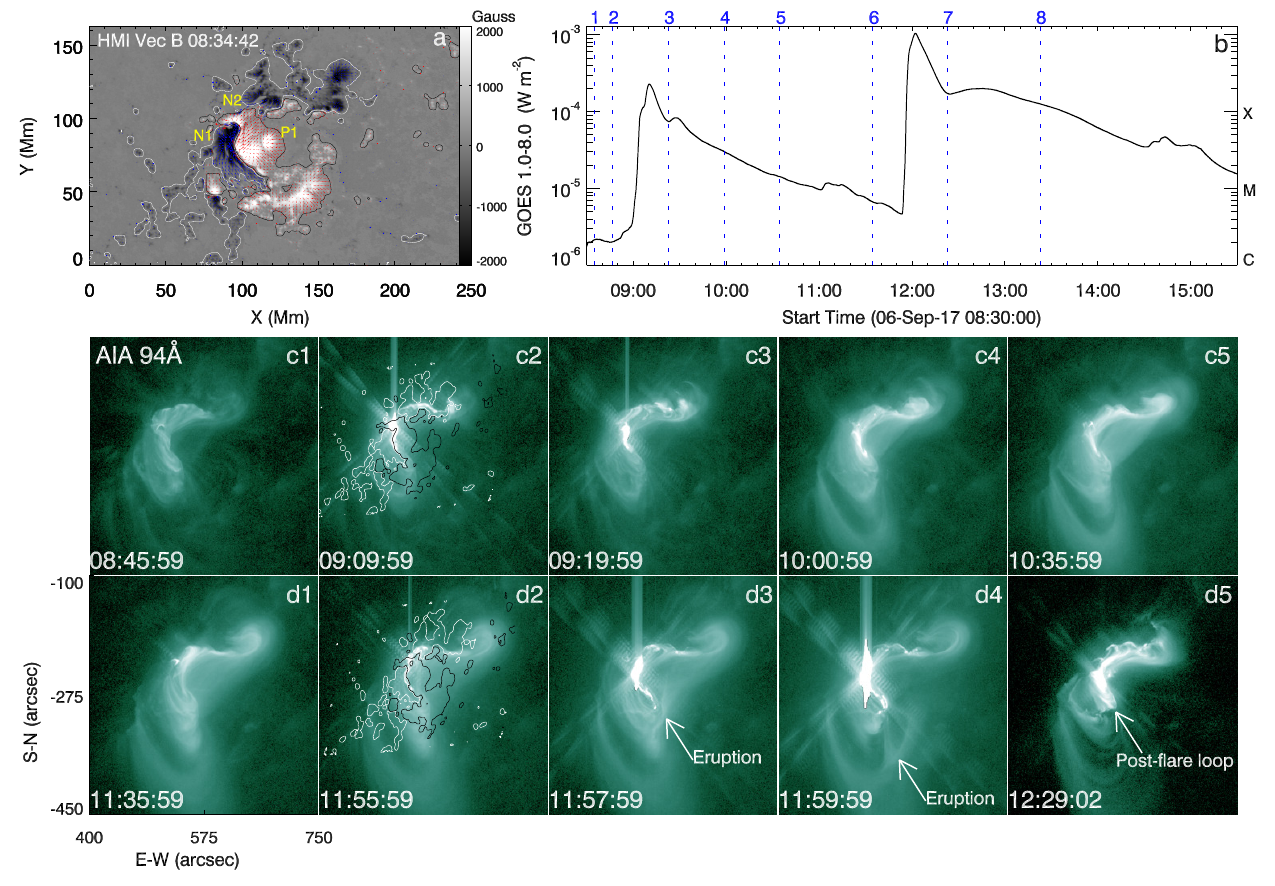}
\caption{{\bf a:} Photospheric 
magnetic fields of the AR. The background shows the vertical fields ($B_z$). 
N1 (P1, N2) refers to the main negative (main positive, northern negative) polarity.  
White (black) contours outline $B_z$ at -150 (150)~Gauss (same in {\bf c2} and {\bf d2}). The arrows refer to the horizontal fields ($B_h$). 
{\bf b:} GOES 1--8 \AA~flux. 
The blue-dashed lines {\bf 1--8} indicate the instances used in Section~\ref{subsec:build_mfr}. 
{\bf c1--c5} ({\bf d1--d5}): Snapshots of the first (second) flare observed in 94~\AA.}\label{fig:obs}
\end{center}
\end{figure*}

NOAA AR 12673 exhibited the fastest magnetic flux emergence ever observed at its early stage~\citep{Sunxd_rna_2017}, 
and had been relatively well-developed when producing the two flares (Figure~\ref{fig:obs}{\bf a}). 
The source of both flares 
involved the central main polarities  
(N1 and P1) and the northern negative polarity (N2). 
Before both flares, hot channels (HCs), which refer to the S-shaped structures appearing above the PILs in high-temperature AIA passbands (94 and 131 \AA) and considered as the proxies of MFRs~\citep[e.g.,][]{Zhangj_2012}, were visible in 94~\AA~(Figure~\ref{fig:obs}{\bf c1} and {\bf d1}). Although the one before the first flare was slightly diffuse, the HCs indicated the existence of MFRs. 
During the first flare, 
brightenings occurred along the HC, 
after which an HC 
was still found (Figure~\ref{fig:obs}{\bf c2--c5}). During the second flare, 
the eruption of the HC 
was observed; post-flare loops appeared later 
(Figure~\ref{fig:obs}{\bf d2--d5}). The observations hint that the first (second) flare may be confined (eruptive).  

\begin{figure*}
\begin{center}
\epsscale{0.85}
\plotone{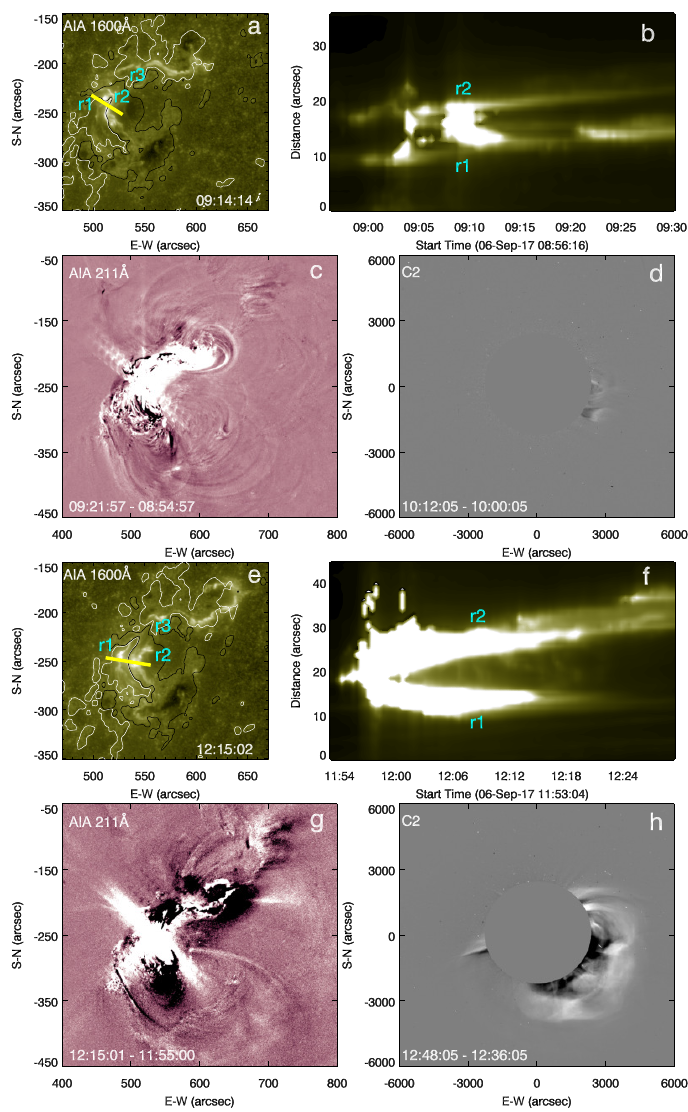}
\caption{{\bf a--d}: Details of the first flare.
{\bf a:} Flare ribbons in 1600~\AA. 
Contours outline $B_z$. 
The letters r1 (r2, r3) denotes the ribbon in the polarity N1 (P1, N2). 
{\bf b:} Time-distance plot of a slice (yellow line in {\bf a}). 
{\bf c:} Base-difference image at the 211~\AA~passband. 
{\bf d:} Coronal outflow observed by LASCO/C2. 
{\bf e--h:} Similar layouts as {\bf a--d} but for the second flare. An animation, including the GOES 1--8 \AA~flux, $B_z$, observations in 1600, 304, 211, 94~\AA~and the coronagraph is available online.}\label{fig:erup}
\end{center}
\end{figure*}

In both flares, three ribbons (r1, r2 and r3 in Figure~\ref{fig:erup}) appeared in 1600~\AA~passband. 
For the first flare, 
the two main 
ribbons (r1 and r2) did not separate as a function of time 
(Figure~\ref{fig:erup}{\bf b}), 
which is in contrast to that of an eruptive flare as stated in standard flare model~\citep[e.g.,][]{Carmichael_1964}, also indicating a confined flare. 
No large-scale dimmings were observed in 211~\AA~(Figure~\ref{fig:erup}{\bf c}).  
Careful inspection of the 
observations in 304~\AA~neither finds ejection signs. 
The associated faint and narrow coronal outflow (Figure~\ref{fig:erup}{\bf d}) 
might not be enough to be defined as a CME. 
For the second flare, 
Figure~\ref{fig:erup}{\bf f} clearly shows that the ribbons separated as the flare evolved. Clear dimmings were seen in 211~\AA~passband after the flare (Figure~\ref{fig:erup}{\bf g}), indicating mass depletion.  
The appeared bright, halo CME (Figure~\ref{fig:erup}{\bf h}) further confirmed the second flare being eruptive. 

\begin{figure*}
\begin{center}
\epsscale{0.95}
\plotone{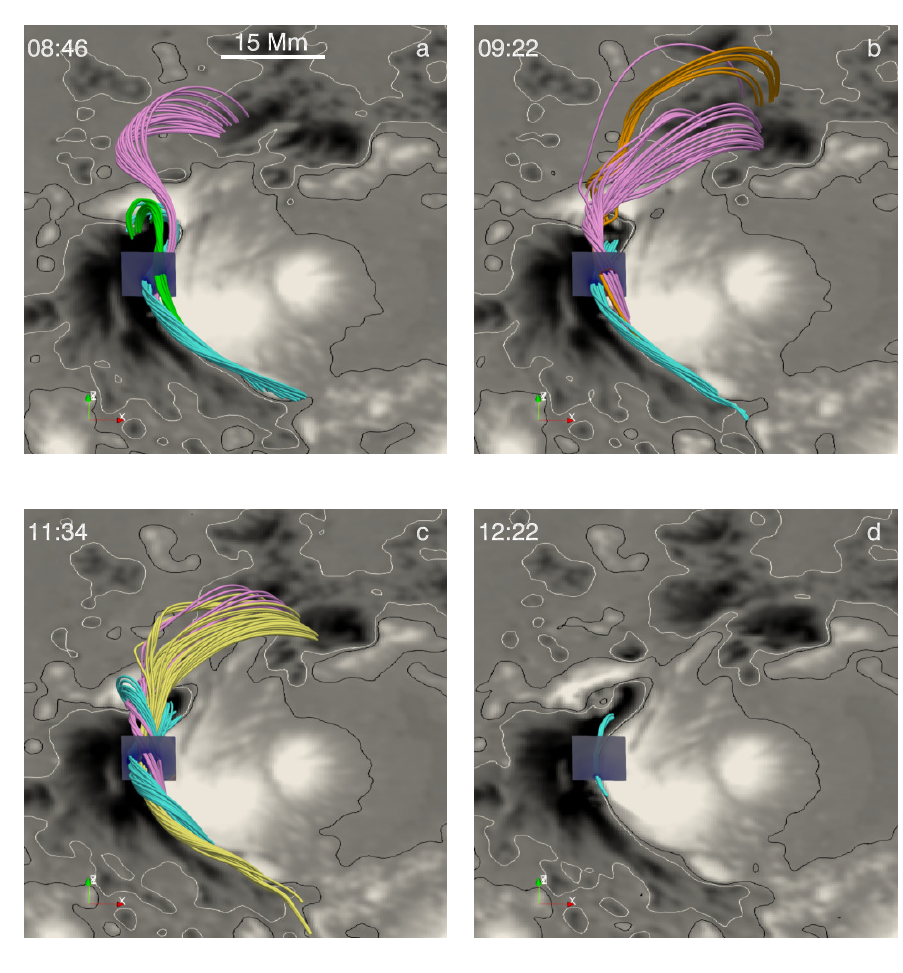}
\caption{{\bf a--b} ({\bf c--d}){\bf:} Representative field lines of the core structure above the PIL 
before and after the confined (eruptive) flare, 
showing an MFR-system in {\bf a--c} and a sheared structure in {\bf d}. 
Different colors represent different branches. The backgrounds are $B_z$. 
The vertical planes mark the position of the plane used in Figure~\ref{fig:fqt}.}\label{fig:ropes} 
\end{center}
\end{figure*}

We calculate $T_w$ and $Q$ in multiple vertical planes across the main PIL to trace the possible MFR before and after the flares (see details in Section~\ref{subsec:build_mfr}). 
The selected field lines 
tracing through twisted/highly sheared regions enclosed by high $Q$ lines are shown in Figure~\ref{fig:ropes}. 
Before the confined flare, 
an MFR-like structure composed of different branches 
of field lines was located (Figure~\ref{fig:ropes}{\bf a}). 
Two branches connected N1 and P1 (cyan and green lines), 
having distinguishable boundaries in the $T_w$ and $Q$ maps (see Section~\ref{subsec:build_mfr}). 
One branch (magenta lines) 
connected P1 and N2. 
The footpoints of the field lines corresponded to the positions of the flare ribbons. Note that, a mature MFR requires its field lines having a coherent structure, which wasn't perfectly met here. 
However, the branches 
were located close, 
evolved as a whole and erupted together finally (see contents below). 
Thus, we name the structure an MFR-system. 
After the confined flare, 
an MFR-system was also found (Figure~\ref{fig:ropes}{\bf b}), 
although its configuration was significantly different from the one before the flare. 
It was still composed of different branches, 
with one branch connecting N1 and P1 (cyan lines), 
two branches connecting P1 and N2 with opposite handedness (magenta and brown lines in Figure~\ref{fig:ropes}{\bf b}, see section~\ref{subsec:build_mfr}). 
The survival of the MFR-system was consistent with the absence of a CME. 

Before the eruptive flare, an MFR-system was still located (Figure~\ref{fig:ropes}{\bf c}), 
though its topology had a slight change, 
appearing longer field lines (shown in yellow). 
The footpoints of these field lines were consistent with the positions of the flare ribbons. 
After the flare, the MFR-system disappeared, 
only a sheared structure survived (Figure~\ref{fig:ropes}{\bf d}). 

In summary, for the X2.2 flare, 
no clear EUV dimmings or separation of its flare ribbons were observed; the accompanied coronal outflow was faint and narrow, indicating a confined flare.
For the X9.3 flare, 
clear on-disk eruption signs, EUV dimmings, separation of its flare ribbons, and the accompanied major CME indicated an eruptive flare. 
The survival and disappearance of the MFR-system after respective flare were consistent with their eruptiveness.

\subsection{Buildup of the MFR-system during the Confined Flare}\label{subsec:build_mfr}

\subsubsection{Boundary of the MFR-system}

To quantitatively study the properties of the MFR-system, 
we select eight instances (vertical-dashed lines in Figure~\ref{fig:obs}{\bf b}), 
at which $T_w$ and $Q$ distributions in a vertical plane are displayed (Figure~\ref{fig:fqt}). We choose the plane 
because it intersected all rope branches. 
The computational grid of the plane is refined by 16 times for a higher precision \citep[following][]{rliu_2016}.  
To additionally calculate $\Phi_t$, $<|T_w|>$, and $H_{twist}$ of the MFR-system, 
its boundary needs to be determined. 
Owing to that perfectly closed QSLs enclosing the MFR-system cannot be found here, the boundaries (white dotted-lines in Figure~\ref{fig:fqt}) 
are determined manually through combining the high-$Q$ lines and contours of the thresholds of $|T_w|$. The thresholds 
are decided when the high $Q$ 
and contour lines connect smoothly, 
resulting in a value of 1.1 before the eruptive flare ({\bf 1-6} in Figure~\ref{fig:fqt}), and 0.7 after the eruptive flare ({\bf{7-8}} in Figure~\ref{fig:fqt}).  
The parameters 
calculated in the 
cross sections are 
shown in Figure~\ref{fig:flux_tw_dc}. 
The fractional changes of $\Phi_t$ and $H_{twist}$ are calculated by $\displaystyle R_{\Phi}=\frac{\Phi_{t}(i)-\Phi_{t}(i-1)}{\Phi_{t}(i-1)}$ and $\displaystyle R_{H}=\frac{H_{twist}(i)-H_{twist}(i-1)}{H_{twist}(i-1)}$, respectively, $i$ means the $i_{th}$ instance.

Uncertainty for each parameter is considered from two sources: 
the selection of the plane and the determination of the MFR-system boundaries. 
For the former, 
we choose another two planes that also intersected all rope branches and repeat the calculation, 
then regard 
the standard deviation of the values from the three planes as 
part of the upper/lower error 
shown in Figure~\ref{fig:flux_tw_dc}. 
For the later, 
we apply the morphological erosion and dilation, with a circular kernel ($r\approx 0.1$~Mm), to the regions in the determined boundaries,
resulting in a shrunken and a distend regions 
where we calculate the parameter again; the difference between the values from the original and the distend (shrunken) regions is regarded as the other part of the upper (lower) error. 
The final errors shown in Figure~\ref{fig:flux_tw_dc} 
are the sum of the two parts.

\subsubsection{Evolution of the Properties of the MFR-system}

\begin{figure*}
\begin{center}
\epsscale{1.1}
\plotone{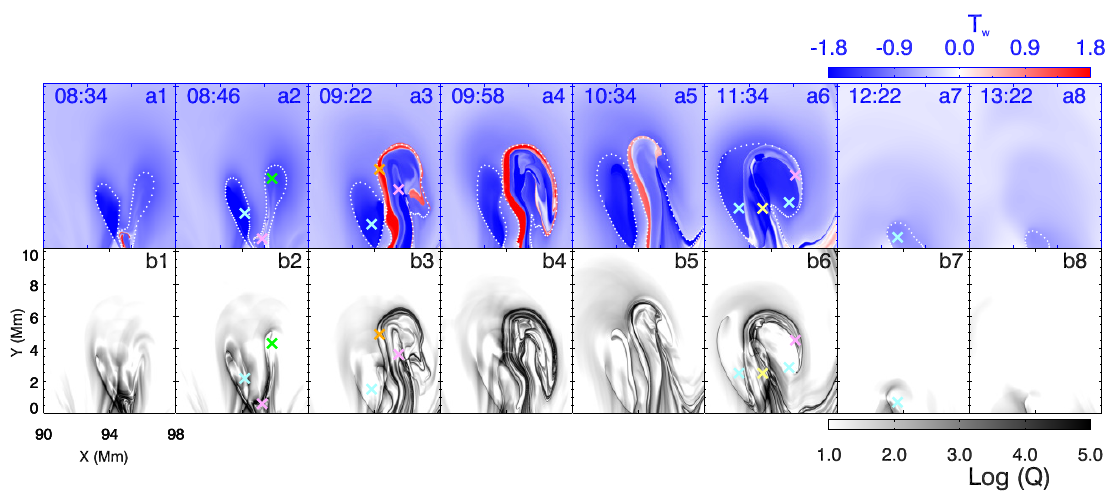}
\caption{The upper (lower) row: 
$T_w$ ($Q$) maps in a plane across the core structure, 
which is an MFR-system in {\bf 1--6} and a sheared structure in {\bf 7--8}.  
Panels {\bf 1--8} correspond to the instances {\bf 1--8} marked in Figure~\ref{fig:obs}{\bf b}.  
$\times$ symbols in {\bf 2}, {\bf 3}, {\bf 6} and {\bf 7} mark the representative positions of different rope branches, 
corresponding to the field lines in Figure~\ref{fig:ropes}. 
The white dots in the upper row outline the boundaries of the core structure. 
}\label{fig:fqt} 
\end{center}
\end{figure*}

At the two instances ({\bf 1} and {\bf 2} in Figure~\ref{fig:fqt}) before the confined flare, 
$T_w$ and $Q$ maps 
showed no significant change. 
Three twisted regions ($\times$ symbols in Figure~\ref{fig:fqt}{\bf a2}, 4{\bf b2}) with distinguishable $Q$ boundaries were located, 
confirming the existence of an MFR-system with different branches, 
as shown in Figure~\ref{fig:ropes}{\bf a}. After the first flare, 
twisted regions with high $Q$ boundaries can still be found ({\bf 3} in Figure~\ref{fig:fqt}), 
consistent with the MFR-system shown in Figure~\ref{fig:ropes}{\bf b}. The right part of the twisted regions showed a field lines set with left handedness (negative twist) overlain by another set with right handedness (positive twist), 
displaying a complex configuration. 
The cross section of the structure displayed an apparent expansion. 
Its equivalent radius increased from 1.6~Mm to 2.6~Mm, indicating a possible enhancement of the axial flux. 
Meanwhile, its estimated length ($\sim$100~Mm) and height (below 10~Mm) didn't change significantly. 

At the instances between the two flares ({\bf 4}, {\bf 5} and {\bf 6} in Figure~\ref{fig:fqt}), twisted regions with 
high $Q$ lines were all located, 
although with their patterns slightly changing, 
indicating that the MFR-system was evolving slowly. After the eruptive flare ({\bf 7} in Figure~\ref{fig:fqt}), the 
twisted region disappeared, 
left a strongly sheared region (with average $|T_w|$ around 0.8 turns), consistent with the eruption of the observed HC. 
One hour later, a sheared structure was still located, having 
lower $T_w$ ($\sim$ 0.6) and smaller cross section ({\bf 8} in Figure~\ref{fig:fqt}). 
 
\begin{figure*}
\begin{center}
\epsscale{0.95}
\plotone{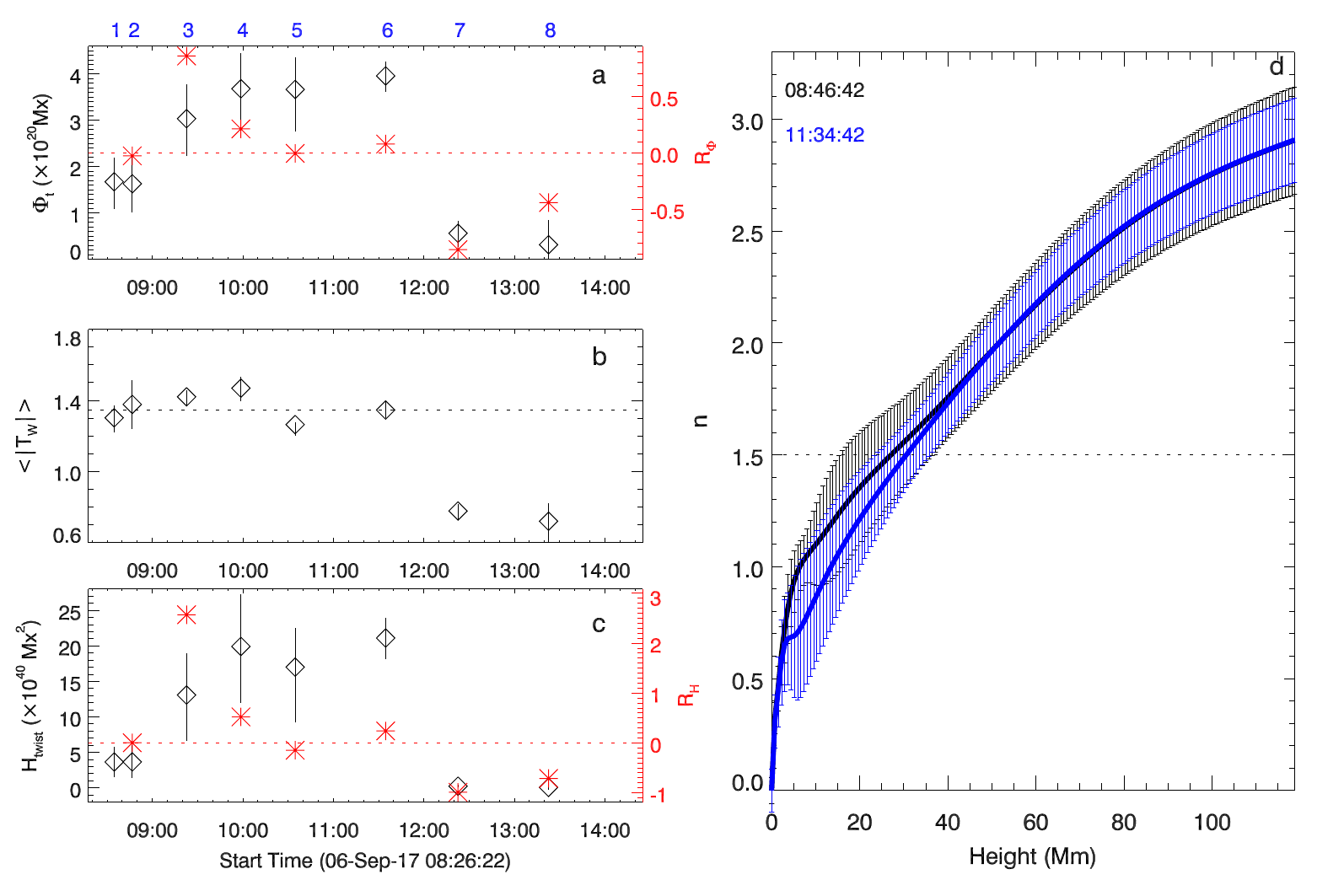}
\caption{{\bf a:} Temporal evolution of $\Phi_t$ 
and its fractional change $R_{\Phi}$. 
{\bf b:} Evolution of $<|T_w|>$. 
The horizontal-dashed line indicates its 
median value (1.35 turns). 
{\bf c:} Evolution of $H_{twist}$. 
{\bf d:} Pre-flare distributions of 
$n$ above the PIL. 
Each data point is the average at this height. The errors are the standard deviations.}\label{fig:flux_tw_dc}
\end{center}
\end{figure*}

The temporal evolution of $\Phi_t$ (Figure~\ref{fig:flux_tw_dc}{\bf a}) showed that it experienced a dramatic increase after the confined flare from $1.63\times10^{20}$~Mx to $3.04\times10^{20}$~Mx (by $86\%$), 
then increased slowly to $3.96\times10^{20}$~Mx at 11:34~UT, 
and finally reduced to $0.56\times10^{20}$~Mx after the eruptive flare. In the mean time, 
$<|T_w|>$ kept a value around 1.3 and dropped to 0.8 before and after the eruptive flare, respectively (Figure~\ref{fig:flux_tw_dc}{\bf b}). 
$H_{twist}$ also dramatically increased from $3.7\times10^{40}$~Mx$^2$ to $13.1\times10^{40}$~Mx$^2$ (by $260\%$) 
after the confined flare, then slowly evolved to $21.1\times10^{40}$~Mx$^2$ at 11:34~UT, 
and decreased to $0.2\times10^{40}$~Mx$^2$ after the eruptive flare (Figure~\ref{fig:flux_tw_dc}{\bf c}). 
Moreover, as the 
decay index distributions in Figure~\ref{fig:flux_tw_dc}{\bf d} showed, before both flares, $n$ reached the critical height (where $n = 1.5$) around 30~Mm, showing no significant difference when considering the computing errors. 

The magnetic flux and 
helicity of the entire AR were also estimated. The former was on the order of $\sim 10^{22}$~Mx. The latter was 
an accumulated value since the emergence of the AR, 
and was roughly assessed based on a velocity estimation method called DAVE4VM~\citep{Schuck_2008}; its value was on the order of $\sim 10^{43}$~Mx$^2$. 
The two parameters of the MFR-system only accounted for $\sim 1\%$ of the values of the AR. 

In summary, we calculate $T_w$ and $Q$ in the cross section of the MFR-system at eight instances, 
and analyze the evolution of its $\Phi_t$, $<|T_w|>$, and $H_{twist}$. 
During the confined flare, the twist number had no significant change;  
interestingly, its axial flux and magnetic helicity were enhanced dramatically. 
After the eruptive flare, the parameters were all significantly reduced. 
The results indicate a rapid buildup and a release process of the MFR-system during the confined and 
eruptive flare, respectively. 
The similar pre-flare distributions of the decay index 
further suggest that the change of the MFR-system itself may play a more significant role in its final 
eruption here. 
 
\section{Summary and Discussions}\label{sec:discuss}

Here we present a buildup process of an MFR-system during a confined X2.2 flare occurred on 2017 September 6 in NOAA AR 12673, 
which was evidenced by significant enhancement of its axial flux and magnetic helicity. 
The MFR-system erupted to a major CME during the following X9.3 flare three hours later. 
 
Note that, the coronal outflow associated with the first flare is classified as a CME in some work~\citep[e.g.,][]{Yanxl_2018a}. 
However, it was very faint, narrow, and did not propagate further than 10~$R_{\odot}$, 
being clearly different from the major CME associated with the second flare. 
This kind of outflows, showing no clear shape of MFRs, failing to propagate to a large distance 
that seem not to be related to MFRs, are defined as ``pseudo-'' CMEs~\citep{Vourlidas_2010, Vourlidas_2012}. The outflow here should not be relevant to the MFR-system we discussed. 
Combined with the evidence in Section~\ref{subsec:obs}, we argue that the X2.2 flare was confined. 
\citet{Verma_2018} also found localized, confined flare kernels in white-light emission of the X2.2 flare and separating ribbons in that of the X9.3 flare, 
consistent with our arguments. 

The structure composed of multiple field lines branches 
here was defined as an MFR-system. 
Note that, we cannot exclude the possibility that those branches were different MFRs 
due to the distinguishable boundaries 
they had, and the opposite helicity signs that different branches further displayed after the first flare. 
However, 
the branches were located close; more importantly, they evolved and erupted together, 
indicating that defining the structure as an MFR-system should be reasonable. 
The QSL of the MFR-system was not perfectly closed, indicating a possible bald patch topology~\citep[e.g.,][]{Savcheva_2012}, and/or ongoing development. 
The evolution of $Q$ and $T_w$ patterns supported its development. When no eruption occurred, the slow evolution of the MFR-system may be driven by sunspot rotation and shear motion~\citep{Yanxl_2018a}. The multiple-branch configuration of an MFR-system has 
been reported~(\citealp[e.g.,][]{Inoue_2016},~\citealp[or double-decker MFR in e.g.,][]{Cheng_2014a}), and is thought to play a role in an eruption when 
interaction exists between different branches~\citep{Awasthi_2018}.
  
Besides the 
final eruption, the most dramatic change of the MFR-system happened during the confined flare. 
Its axial flux and magnetic helicity increased by $86\%$ and $260\%$ within tens of minutes, 
clearly evidencing a rapid buildup of the MFR-system. 
The process can be achieved by the fast flare reconnection, which can 
add flux to a pre-existing MFR efficiently. 
Here the reconnection may start at the QSL between different branches, 
involving not only the rope fields but also the surrounding fields, adding more flux to the rope. 
Correspondence between the footpoints of the MFR-system and 
the flare ribbons also supported such a scenario. 
The mean twist number did not significantly change during the confined flare, 
indicating that the toroidal and poloidal flux were 
added to the MFR-system with equal proportion, 
keeping the twist number relatively invariant. 
Meanwhile, the 
decay index distributions showed no significant evolution.  
The result was in agreement with~\citet{Nindos_etal_2012}, 
in which small temporal evolution of the decay index, accompanied by helicity buildup, was found 
during a multi-day period when various 
eruptive and non-eruptive flares occurred.

The increase of the MFR-system flux may lead to a easier catastrophe~\citep[e.g.,][]{Zhangqh_2016}. 
The enhanced magnetic helicity of the MFR-system is related to the helicity of the current-carrying magnetic fields, 
the ratio of which to the total helicity is positively correlated to the CME eruptivity of a region~\citep{Pariat_2017}. 
Based on the fact of nearly invariant 
decay index, 
we conjecture that the first confined flare enhanced the MFR-system so that facilitated the next successful eruption. The result is also consistent 
with a so-called ``domino-effect'' scenario~\citep{Zuccarello_2009}, 
which emphasizes the influence from the previous activity. 
One should pay more attention to the close precursors of the eruptions~\citep{Wangym_2013e, Lliu_2017}, 
since they might play roles in facilitating the final eruptions. 

\acknowledgments{We thank our anonymous referee for his/her constructive comments that significantly improved the manuscript. We thank Prof. Jie Zhang for his helpful discussions. 
We acknowledge 
the data from {\it{GOES}}, {\it SDO}, and {\it SOHO}. 
L.L. is supported by the Open Project of CAS Key Laboratory of Geospace Environment, and NSFC (11803096). 
X.C. is funded by NSFC (11722325, 11733003, 11790303, 11790300), Jiangsu NSF (BK20170011), and “Dengfeng B” program of Nanjing University. 
Y.W. is supported by NSFC (41574165, 41774178). 
Y.G. is supported by NSFC (11773016, 11533005) and the fundamental research funds for the central universities 020114380028.
J.C. is supported  by NSFC (41525015, 41774186).

\clearpage
\bibliographystyle{aasjournal} 

\end{document}